\documentclass[twocolumn,aps,prl]{revtex4}
\usepackage{graphicx}

\begin{document}
\bibliographystyle{prsty}
\title{Observation of Competing Order in a High-$T_{c}$
Superconductor with Femtosecond Optical Pulses}
\author{Elbert E. M. Chia}
\author{Jian-Xin Zhu}
\author{D. Talbayev}
\author{R. D. Averitt}
\altaffiliation{Department of Physics, Boston University, 590
Commonwealth Avenue, Boston MA 02215, USA}
\author{A. J. Taylor}
\affiliation{Los Alamos National Laboratory, Los Alamos, NM 87545,
USA}
\author{Kyu-Hwan Oh}
\author{In-Sun Jo}
\author{S.-I. Lee}
\altaffiliation{Quantum Materials Research Laboratory, Korea Basic
Science Institute, Daejeon 305-333, Republic of Korea}
\affiliation{National Creative Research Initiative Center for
Superconductivity and Department of Physics, Pohang University of
Science and Technology, Pohang 790-784, Republic of Korea}
\date{\today}

\begin{abstract}
We present studies of the photoexcited quasiparticle dynamics in
Tl$_{2}$Ba$_{2}$Ca$_{2}$Cu$_{3}$O$_{y}$ (Tl-2223) using femtosecond
optical techniques. Deep into the superconducting state (below 40
K), a dramatic change occurs in the temporal dynamics associated
with photoexcited quasiparticles rejoining the condensate. This is
suggestive of entry into a coexistence phase which, as our analysis
reveals, opens a gap in the density of states (in addition to the
superconducting gap), and furthermore, competes with
superconductivity resulting in a depression of the superconducting
gap.
\end{abstract}

\maketitle In the Bardeen-Cooper-Schrieffer (BCS) theory of
superconductivity, which describes the mechanism of conventional
superconductivity for conventional metals, electrons form Cooper
pairs mediated by the vibrations of the crystal lattice. For the
high-temperature superconductors, another possibility exists,
namely, Cooper pairing via antiferromagnetic spin fluctuations
\cite{Zhang97,Arovas97}. Indeed, a full-fledged antiferromagnetic
order, out of which such antiferromagnetic fluctuations emerge, can
also compete with superconductivity as the dominant ground state
resulting in phase coexistence
\cite{Lee99,Pan01,Kotegawa04,Lake02,Mukuda06}. The coexistence of
antiferromagnetic ordering with superconductivity has been observed
in single- or double- layer systems in the presence of a magnetic
field via neutron scattering \cite{Lake02,Kang03}, or in
five-layered systems in zero field using nuclear magnetic resonance
\cite{Kotegawa04,Mukuda06}. However, it is unclear from these
measurements how the emergence of antiferromagnetic order affects
the quasiparticle (QP) excitations, which determine the material's
optical and electronic response.

An energy gap in the excitation spectrum of a material is of
fundamental importance in determining its optical and electronic
properties. In superconductors and many other correlated electron
materials, many-body interactions open a gap in the QP density of
states thereby introducing an additional timescale for the QP
dynamics. Independent of its origin, the opening of a gap presents a
bottleneck to ground state recovery following photoexcitation of QPs
across the gap. The timescale of this recovery is related to the gap
magnitude, meaning that interactions which perturb the gap manifest
as an easily measured change in the temporal response by monitoring
changes in the reflectivity ($\Delta R/R$) or transmission of an
interrogating probe beam. In recent years, femtosecond time-resolved
spectroscopy has been recognized a powerful \textit{bulk} technique
to study temperature-dependent changes of the low-lying electronic
structure of superconductors \cite{Han90} and other strongly
correlated electron materials \cite{Kabanov00,Averitt01,Demsar03}.
It provides a new avenue, namely the time domain, for understanding
the QP excitations of a material. In this Letter, we present
time-resolved studies of photoexcited QP dynamics in the
high-$T_{c}$ superconductor Tl-2223. We observe that its pristine
superconducting state (40~K $<$ $T$ $<$ $T_{c}$) subsequently
evolves into a coexistence phase as evidenced by a strong
modification of the gap dynamics below 40~K.

Our sample is a slightly underdoped single crystal of Tl-2223 with
$T_{c}$=115~K, grown by the self-flux method \cite{Kim04}. Tl-2223
is a tri-layered crystal, where its two outer CuO$_{2}$ planes has a
pyramidal coordination with an apical oxygen, while the inner plane
has a square coordination with no apical oxygen. In our experiments
an 80-MHz repetition rate Ti:sapphire laser produces 80-fs pulses at
$\approx$800~nm (1.5~eV) as the source of both pump and probe
optical pulses. The pump and probe pulses were cross-polarized, with
a pump spot diameter of 60~$\mu$m and probe spot diameter of
30~$\mu$m. The reflected probe beam was focused onto an avalanche
photodiode detector. The pump beam was modulated at 1~MHz with an
acoustic-optical modulator to minimize noise. The average pump power
was 300~$\mu$W, giving a pump fluence of $\sim$0.01 $\mu$J/cm$^{2}$
and a photoexcited QP density of 0.002/unit cell, showing that the
system is in the weak perturbation limit. The probe intensity was 10
times lower. Data were taken from 7~K to 300~K. The photoinduced
temperature rise at the lowest temperatures was estimated to be
$\sim$10~K (in all the data the temperature increase of the
illuminated spot has been accounted for). The small spot sizes
enable us to use a very small pump fluence without sacrificing
signal-to-noise ratio, while keeping sample heating to a minimum and
thus enabling data to be taken at low temperatures. The resolution
is at least 1 part in 10$^{6}$. We also used a 2-MHz repetition rate
cavity-dumped laser to obtain data at 7~K.

\begin{figure} \centering \includegraphics[width=8cm,clip]{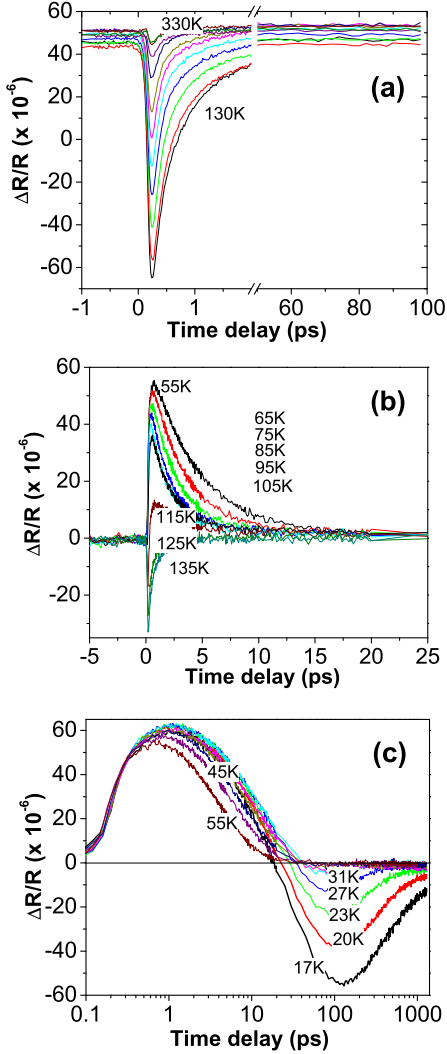}
\caption{Photoinduced transient reflection $\Delta R/R$ versus time
delay between pump and probe pulses, at a series of temperatures
through $T_{\varphi}$ and $T_{c}$. The logarithmic scale is used for
the time-axis in the low-temperature region (Fig. 1c).}
\label{fig:Tl2223Fig1}
\end{figure}

\begin{figure}
\centering \includegraphics[width=8cm,clip]{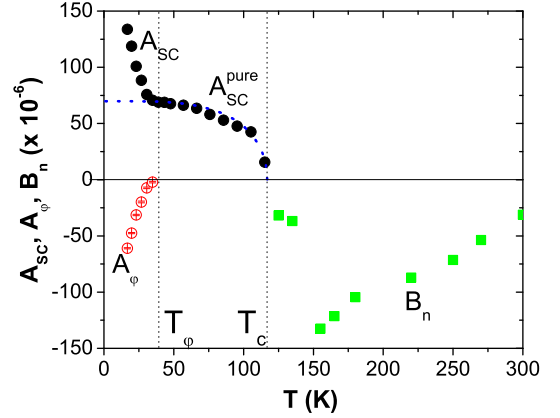}
\caption{Temperature dependence of the peak amplitudes. Solid
circles: $A_{SC}$. (O): $A_{\varphi}$. Solid squares: $B_{n}$.
Dotted line: BCS fit to $A_{SC}$ in the temperature range
$T_{\varphi} < T < T_{c}$, extrapolated to $T$=0.}
\label{fig:Tl2223Fig2}
\end{figure}

Figure~\ref{fig:Tl2223Fig1} shows the time dependence of the
photoinduced signal of Tl-2223. At high temperatures the signal is
characterized by a negative $\Delta R/R$ transient which relaxes
within $\tau_{n}$$\sim$0.5 ps with $\tau_{n}$ decreasing slightly as
temperature is increased to 300~K (Fig.~\ref{fig:Tl2223Fig1}a)
consistent with QP thermalization in conventional metals
\cite{Groenveld95}. Below $T_{c}$, we observe the onset of a
positive $\Delta R/R$ with a relaxation time ($\tau_{SC}$) of a few
picoseconds due to the opening of the superconducting gap
(Fig.~\ref{fig:Tl2223Fig1}b). Surprisingly, below $\sim$40~K,
$\Delta R/R$ first goes positive, relaxes to zero with a lifetime
$\tau_{SC}$, \textit{then crosses zero and goes negative}, before
relaxing back to equilibrium over a time scale of a few hundred
picoseconds (Fig.~\ref{fig:Tl2223Fig1}c). We ascribe the short-decay
positive signal to the reformation of superconducting order
following photoexcitation, and the \textit{long-decay negative
signal to the development of a new competing order other than
superconductivity}. We define the second transition temperature as
$T_{\varphi}$. Accordingly, we fit the data of $\Delta R/R$ in
different temperature ranges as follows: In the normal state ($T >
T_{c}$), the data follow $\Delta R/R = B_{0} + B_{n}
\exp(-t/\tau_{n}$), where $B_{n} < 0$; in the phase with only the
superconducting order parameter ($T_{\varphi} < T < T_{c}$), the
data follow $\Delta R/R= A_{0} + A_{SC} \exp(-t/\tau_{SC}$), where
$A_{SC}
> 0$; in the coexistence region, ($T < T_{\varphi}$), the data
follow $\Delta R/R = A_{0} + A_{SC} \exp(-t/\tau_{SC}) + A_{\varphi}
\exp(-t/\tau_{\varphi})$, where $A_{SC} > 0$ and $A_{\varphi} < 0$.
Figure~\ref{fig:Tl2223Fig2} shows the temperature-dependence of the
peak amplitudes $A_{SC}(T)$, $A_{\varphi}(T)$ and $B_{n}(T)$. We see
that below $\sim$40~K, $A_{\varphi}$ increases from zero, while
$A_{SC}$ exhibits a sharp kink.

We use the Rothwarf-Taylor (RT) model to explain our data
\cite{Rothwarf67}. It is a phenomenological model used to describe
the relaxation of photoexcited superconductors, where the presence
of a gap in the electronic density-of-states gives rise to a
bottleneck for carrier relaxation. When two QPs with energies $\ge
\Delta$ recombine ($\Delta$ is the superconducting gap magnitude), a
high-frequency boson (HFB) with energy $\omega \ge 2\Delta$ is
created. The HFBs that remain in the excitation volume can
subsequently break additional Cooper pairs effectively inhibiting QP
recombination. Superconductivity recovery is governed by the decay
of the HFB population. The RT analysis for a material with a single
energy gap, is as follows \cite{Demsar06}: from the
temperature-dependence of the amplitude $A$, one obtains the density
of thermally excited QPs $n_{T}$ via $n_{T} \propto A^{-1} -1$,
where $A(T)$ is the normalized amplitude [$A(T) = A(T)/A(T
\rightarrow 0)$]. Then we can fit the $n_{T}$-data to the QP density
per unit cell
\begin{equation}
n_{T} \propto \sqrt{\Delta (T) T} \exp (-\Delta (T)/T),
\label{eqn:nT}
\end{equation} with $\Delta (0)$ as a fitting parameter and $\Delta (T)$
obeying a BCS temperature dependence. Moreover, for a constant pump
intensity, the temperature dependence of $n_{T}$ also governs the
temperature-dependence of the relaxation time $\tau$, given by
\cite{Demsar06,Chia06}
\begin{equation}
\tau^{-1}(T) = \Gamma [\delta + 2 n_{T} (T)](\Delta + \alpha T
\Delta^{4}), \label{eqn:tau}
\end{equation} where $\Gamma$, $\delta$ and $\alpha$ are
temperature-independent fitting parameters, with $\alpha$ having an
upper limit of $52/(\theta_{D}^{3}T_{min})$, $\theta_{D}$ being the
Debye temperature and $T_{min}$ the minimum temperature of the
experiment.

\begin{figure}
\centering \includegraphics[width=9cm,clip]{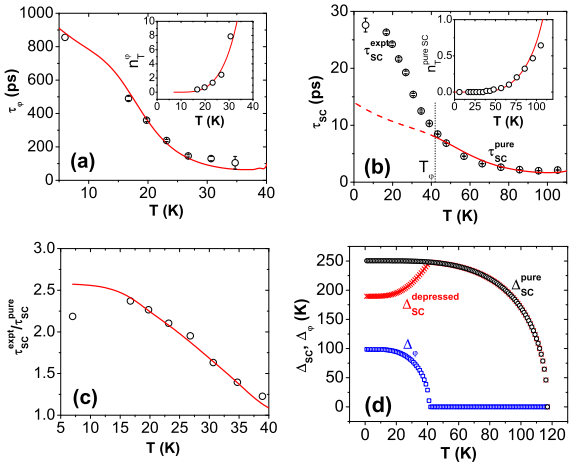}
\caption{Temperature dependence of relaxation times $\tau$ and
thermally-excited QP densities $n_{T}$. (a) $\tau_{\varphi}(T)$ and
$n_{T}(T)$ of the competing order component. Solid curves are fits
using the RT model. (b) $\tau_{SC}(T)$ and $n_{T}(T)$ of the
superconducting component. Solid curves are fits using the RT model.
The RT model fit to $\tau_{SC}(T)$ is only for $T_{\varphi} < T <
T_{c}$ (solid line). The dashed line is the extrapolation of that
fit below $T_{\varphi}$. (c) Experimental values (O) and theoretical
fit (solid line) of the ratio $\tau_{SC}^{expt}/\tau_{SC}^{pure}$.
(d) Superconducting gap $\Delta_{SC}(T)$ with and without
suppression, and the gap $\Delta_{\varphi}(T)$ due to the competing
order, using the Ginzburg-Landau theory.} \label{fig:Tl2223Fig3}
\end{figure}

We first consider the competing component below $T_{\varphi}$. In
this regime we have two types of ordering --- the competing and
superconducting order. Since the nature of the competing order is
unknown \textit{a priori}, we assume that its formation opens an
isotropic QP gap (we discuss the microscopic nature of the competing
order below). Therefore, the bottleneck effect is mostly dominated
by this new gap. We then fit the data for the new order
($A_{\varphi}$ and $\tau_{\varphi}$) using the RT model described
above. That is, from $A_{\varphi} (0)$ and $A_{\varphi} (T)$, we
obtain $n_{T}$ (circles in the inset of Fig.~\ref{fig:Tl2223Fig3}a).
Then we fit $n_{T}(T)$ with Eq.~\ref{eqn:nT} (solid line in the
inset of Fig.~\ref{fig:Tl2223Fig3}a) using the gap values
$\Delta_{\varphi} (T)$ obtained from Ginzburg-Landau theory shown in
Fig.~\ref{fig:Tl2223Fig3}d. Finally, we insert the fitted values of
$n_{T}(T)$ into Eq.~\ref{eqn:tau} to determine the experimental
values of $\tau_{\varphi}$, as shown in Fig.~\ref{fig:Tl2223Fig3}a.
The excellent fit lends strong support to our assumption of the
opening of a QP gap upon the development of the competing order. The
dynamics of this competing order can be explained by a relaxation
bottleneck associated with the presence of a gap in the density of
states.

Next we turn to the superconducting component. In the range
$T_{\varphi} < T < T_{c}$, with only one order parameter in the
system, namely superconducting order, we fit the data to the
single-component RT model described above, where recombination
occurs only from the superconducting energy gap. The signal
amplitude is labelled $A_{SC}^{pure} (T)$, with the superscript
denoting the pure superconducting component without the existence of
a competing order. $A_{SC}^{pure} (T=0)$ is not the experimental
value shown on Fig.~\ref{fig:Tl2223Fig2}, but is that extrapolated
from $A_{SC}^{pure} (T > T_{\varphi})$, via the BCS temperature
dependence, assuming that the competing order does not exist. This
is shown as a dashed line on Fig.~\ref{fig:Tl2223Fig2}, whence one
obtains $A_{SC}^{pure} (T=0) \approx$ 70.0 x 10$^{-6}$. From
$A_{SC}^{pure} (0 < T < T_{c})$ one then obtains $n_{T}^{pureSC}(T)$
as shown in the inset of Fig.~\ref{fig:Tl2223Fig3}b (circles), where
the fit to Eq.~\ref{eqn:nT} yields $\Delta (0)$ = 2.14k$_{B}T_{c}$
(solid line), in agreement with the typical $d$-wave value. Again,
using these fitted values of $n_{T}^{pureSC} (T)$, one fits the
experimental values of $\tau_{SC}^{pure} (T)$ in the range
$T_{\varphi} < T < T_{c}$ using Eq.~\ref{eqn:tau}, shown on
Fig.~\ref{fig:Tl2223Fig3}b. Similar to the competing phase above,
the relaxation dynamics of the pure superconducting phase can also
be explained by the presence of a relaxation bottleneck due to a
(superconducting) gap in the density of states.

We immediately notice from Fig.~\ref{fig:Tl2223Fig3}b that below
$T_{\varphi}$, the fitted values $\tau_{SC}^{pure} (T)$ (dashed
line) underestimate the experimental values. In the Ginzburg-Landau
theory \cite{Zhu01}, the coupling between the competing and
superconducting order parameters causes the superconducting gap to
be suppressed. Hence the superconducting energy gap $\Delta_{SC}$
decreases below its BCS value, as shown in
Fig.~\ref{fig:Tl2223Fig3}d. Since at a fixed temperature, $\tau$
increases as $\Delta$ decreases (and vice versa) \cite{Kabanov00},
we can infer that, below $T_{\varphi}$, the increase of the
experimental relaxation time $\tau_{SC}^{expt} (T)$ over its BCS
value $\tau_{SC}^{pure} (T)$ is due to the suppression of the
superconducting gap in this temperature range. From
Eq.~\ref{eqn:tau}, we can more accurately put
\begin{equation}
\frac{\tau_{SC}^{expt}}{\tau_{SC}^{pure}} \propto
\frac{\Delta_{SC}^{pure} + \alpha T
(\Delta_{SC}^{pure})^{4}}{\Delta_{SC}^{suppressed} + \alpha^{\prime}
T (\Delta_{SC}^{suppressed})^{4}}, \label{eqn:tauratio}
\end{equation} where $\alpha$ is obtained earlier from the fit to $\tau_{SC}^{pure} (T)$,
and $\alpha^{\prime}$ is a fitting parameter.
Figure~\ref{fig:Tl2223Fig3}c shows the ratio
$\tau_{SC}^{expt}/\tau_{SC}^{pure}$ (circles), and the fit given by
Eq.~\ref{eqn:tauratio} (solid line). We see that the fit reproduces
the general shape of $\tau_{SC}^{expt}/\tau_{SC}^{pure}$ --- an
increase below 40~K and flattening out around 15~K. Our analysis
shows that the deviation of $A_{SC}$ and $\tau_{SC}$ from the BCS
temperature dependence below $T_{\varphi}$ is due to the
\textit{depression} of the superconducting gap, which is caused by
the appearance of the second order. It confirms the
\textit{competing} nature of this new order below $T_{\varphi}$.

Our work is unique compared to other techniques in various aspects.
First, we see the coexistence phase in zero field, while neutron
scattering data on La$_{2-x}$Sr$_{x}$CuO$_{4}$ and
Nd$_{1.85}$Ce$_{0.15}$CuO$_{4}$ sees the emergence of the
antiferromagnetic phase only with an externally applied magnetic
field \cite{Lake02,Kang03}. Thus our data are not complicated by the
presence of vortex lattice and/or stripe order. Second, this is the
first observation of the coexistence phase using ultrafast
spectroscopy, a tabletop setup compared to large facilities required
for neutron scattering experiments. Third, our technique only
requires a sample volume of $\sim$10$^{-10}$ cm$^{3}$ (due to a
small laser spot diameter of 60~$\mu$m and skin depth in the
cuprates of $\sim$80~nm), which is orders of magnitude smaller than
that in neutron scattering ($\sim$1~cm$^{3}$). This makes our
technique especially suitable for ultra-thin platelet-like samples
such as the cuprates, enabling us to probe a much wider class of
cuprate superconductors. Fourth, a tri-layered cuprate system has
the highest $T_{c}$ in a homologous series, and as compared to other
members with a higher number of layers, has the closest charge
distribution between the outer and inner CuO$_{2}$ planes and exists
in a single phase \cite{Chen99}. Ours is the first observation of
the coexistence of a competing order with superconductivity in a
tri-layered system. Fifth, we have successfully applied the RT model
to systems with more than one gap in the density of states, where we
\textit{quantified} the reduction of the superconducting gap in the
presence of a competing order. Though our technique cannot determine
whether this new order is magnetic or not, our data clearly show
that it \textit{competes} with superconductivity. The emergence of
this new order opens a QP gap, and our data can be fit excellently
with a BCS-like gap, indicating the new order is not inconsistent
with a commensurate antiferromagnetic spin-density-wave as revealed
in zero-field NMR data on five-layered polycrystalline cuprates
\cite{Kotegawa04,Mukuda06}. However, we do not exclude the
possibility that the competing order can be $d$-density wave order
\cite{Chakravarty01}, circulating current order \cite{Varma99}, or
charge density wave order \cite{Podolsky03}. Contrast this
coexistence phase at zero field in multi-layered samples with the
situation in single-layered cuprates, where at a finite hole doping
only a single superconducting phase exists and the competing phase
must be induced by an external perturbation such as dc magnetic
field. A possible reason is that for multi-layered cuprates, the
competing and superconducting order may nucleate on different
planes, with each of their correlation lengths much larger than the
interlayer distance, such that the two orders can penetrate into
each other even at zero magnetic field. It is precisely the ability
of ultrafast spectroscopy to \textit{temporally} resolve the
dynamics of different degrees of freedom that enables us to observe
these two orders in the coexistence phase.

We took data on another underdoped sample of Tl-2223 with a higher
$T_{c}$ of 117~K, obtaining a lower $T_{\varphi}$ of 35~K. This is
consistent with the phase diagram of multi-layered cuprates as
depicted in Fig.~4 of Ref.~\onlinecite{Mukuda06}, where in the
(underdoped) coexistence region, the antiferromagnetic transition
temperature $T_{N}$ decreases with increasing doping. Moreover, our
recent data on the two-layered cuprate Tl-2212 do not show the zero
crossover. This is consistent with ultrafast relaxation data on
other one- and two-layered cuprates \cite{Han90,Gay99,Smith00} where
the coexistence phase is not expected to exist, showing that our
observation of the zero crossover in Tl-2223 is intrinsic and not an
artifact of our experimental setup.

Our work presents the first ultrafast optical spectroscopy probe of
the coexistence phase in a multi-layered cuprate superconductor
where, in zero magnetic field, a new order competes with
superconductivity. This competing order is intrinsic to the material
and is not induced by any external applied field. The competing
order opens up a QP gap, consistent with a commensurate
antiferromagnetic order. Our study once again points to the unique
characteristic that high-temperature superconductivity results from
the competition between more than one type of order parameter. It
provides an insight into the mechanism of strongly correlated
superconductivity
--- the quantum fluctuations around this competing order might be
responsible for gluing the electrons into Cooper pairs. Theoretical
work to solve the RT model in the presence of multiple gaps, as well
as experimental studies on other multi-layered cuprates, are clearly
needed to elucidate the temporal dynamics of the coexistence phase
in high-$T_{c}$ superconductors.

We acknowledge useful discussions with J. Sarrao, D. Basov, X.-J.
Chen, K. Burch and J. D. Thompson. Work at Los Alamos was supported
by the Los Alamos LDRD program. E.E.M.C. acknowledges G. T. Seaborg
Postdoctoral Fellowship support.

\bibliographystyle{prsty}
\bibliography{Tl2223}
\bigskip

\end{document}